\documentclass[12pt]{iopart}
\usepackage{iopams}
\usepackage{graphicx,psfrag,color}
\usepackage{amssymb,latexsym,mathrsfs,mathrsfs}
\begin{document}
\def\be{\begin{equation}}
\def\ee{\end{equation}}
\def\bea{\begin{eqnarray}}
\def\eea{\end{eqnarray}}
\title{Resetting of fluctuating interfaces at
power-law times}
\author{Shamik Gupta$^1$, Apoorva Nagar$^2$}
\ead{shamikg1@gmail.com,apoorva.nagar@iist.ac.in}
\address{$^1$ \mbox{Max Planck Institute for the Physics of Complex Systems}, Noethnitzer Stra\ss e 38, D-01187 Dresden, Germany \\
$^2$ Indian Institute of Space Science and Technology, Thiruvananthapuram, Kerala, India}
\begin{abstract}
What happens when the time evolution of a fluctuating interface is interrupted with resetting
to a given initial configuration after random time intervals $\tau$ distributed as a power-law $\sim \tau^{-(1+\alpha)};~\alpha > 0$?
For an interface of length $L$ in one dimension, and an initial flat
configuration, we show that depending on $\alpha$, the dynamics in the limit $L \to \infty$ exhibits a spectrum of
rich long-time behavior. It is known that without resetting, the interface width grows unbounded with time as $t^\beta$ in this limit, where $\beta$ is the so-called growth exponent characteristic
of the universality class for a given interface dynamics. We show that 
introducing resetting induces for $\alpha>1$ and at long times
fluctuations that are bounded in time. Corresponding to such a
reset-induced stationary state is a distribution of fluctuations that is strongly non-Gaussian, with tails decaying as a power-law. The
distribution exhibits a distinctive cuspy behavior for small argument,
implying that the stationary state is out of equilibrium. For
$\alpha<1$, on the contrary, resetting to the flat configuration is
unable to counter the otherwise unbounded growth of fluctuations in time, so that the distribution of fluctuations remains 
time dependent with an ever-increasing width even at long times. Although stationary for $\alpha>1$,
the width of the interface grows forever with time as a power-law for
$1<\alpha < \alpha^{({\rm w})}$, and converges to a finite constant only
for larger $\alpha$, thereby exhibiting
a crossover at $\alpha^{({\rm w})}=1+2\beta$. The time-dependent distribution of fluctuations for $\alpha<1$ exhibits for small argument another
interesting crossover behavior, from cusp to divergence, across $\alpha^{({\rm d})}=1-\beta$. We demonstrate these results by exact analytical
results for the paradigmatic Edwards-Wilkinson (EW) dynamical evolution of the interface, and further corroborate our findings by extensive numerical
simulations of interface models in the EW and the Kardar-Parisi-Zhang universality class.
\end{abstract}
\date{\today}
\maketitle
\section{Introduction}
Stochastic processes that incorporate incremental changes in the state of a 
dynamical variable in a small time, interspersed with sudden large
changes occurring at random time intervals, are rather common in nature.
A particular class of a sudden change is a reset to a given state.
Considering simple diffusion as the incremental process, studies of the opposing
effects of diffusive spreading away from a given state and
confinement around the same state due to repeated resetting at random
intervals have been a recurring theme of research in recent years.
A variety of situations have been considered, e.g., a diffusing particle resetting to its initial position in either a free
\cite{Evans:2011-1,Majumdar:2015-1,Reuveni:2015,Eule:2016,Nagar:2015},
or a bounded domain \cite{Christou:2015}, in presence of an external
potential \cite{Pal:2015}, for different choices
of resetting position \cite{Evans:2011-2,Boyer:2014,Majumdar:2015-2}.
Further generalizations that were considered in the literature include resetting of continuous-time random walks
\cite{Montero:2013,Mendez:2016}, L\'{e}vy \cite{Kusmierz:2014} and
exponential constant-speed flights \cite{Campos:2015}, and
time-dependent resetting of a Brownian particle \cite{Pal:2015-1}.
Stochastic resetting has also been invoked in the context of
reaction-diffusion models \cite{Durang:2014}, in backtrack
recovery by RNA polymerases \cite{Roldan:2016}, and even in discussing stochastic thermodynamics far from equilibrium \cite{Fuchs:2016}.
A particular class of systems for which stochastic resetting has been shown
to lead to novel features is that of fluctuating interfaces in one
dimension ($1d$) \cite{Gupta:2014}. 

Examples of fluctuating interfaces abound in nature, e.g., in fluid flow in porous
media, in vortex lines in disordered superconductors, in liquid-crystal
turbulence, in the field of molecular beam epitaxy, in fluctuating steps
on metals, in growing bacterial colonies or tumor, and others
\cite{Barabasi:1995,Halpin-Healy:1995,Krug:1997}.
Fluctuating interfaces constitute an important example of
an extended many-body interacting system with non-trivial correlations
between the constituents. Such correlations strongly affect the nature
of the long-time stationary state (when it exists), and also the relaxation
towards it. An important task in this regard concerns analyzing the
nature of fluctuations for a given dynamics of interface evolution, and identifying the associated universality class
characterized by critical exponents that describe quantitatively various
statistical properties of the fluctuations in the scaling limit
\cite{Krug:1997}. 

A well-studied model of fluctuating
interfaces that allows exact determination of scaling functions and
exponents and unveiling of the non-trivial effects of
correlations is the Edwards-Wilkinson (EW) interface \cite{EW:1982}.
Such an interface describes, e.g., a surface generated by random
deposition of particles onto a substrate followed by their diffusion
along the surface \cite{Barabasi:1995}. To describe the model in $1d$, consider a substrate
of length $L$, and an interface characterized by the height $H(x,t)\ge0$ above
position $x \in [0,L]$ at time $t$. The EW interface
evolves in time according to the linear equation \cite{EW:1982} 
\be
\frac{\partial H}{\partial t}=\nu\frac{\partial^{2}H}{\partial x^{2}}+\eta(x,t),
\label{eq:eom-ew}
\ee
where $\nu$ is the diffusivity, and the Gaussian, white noise
$\eta(x,t)$ satisfies
$\langle\eta(x,t)\rangle=0,\langle\eta(x,t)\eta(x',t')\rangle=2D\delta(x-x')\delta(t-t')$.
Here, angular brackets denote averaging over noise, while $D$
characterizes the strength of the noise. It is usual to start with a
flat interface: $H(x,0)=0~\forall~x$, and, additionally, consider periodic boundary conditions:
$H(0,t)=H(L,t) ~\forall~ t$.

Denoting by $\overline{H(x,t)}\equiv(1/L)\int_{0}^{L}dx~ H(x,t)$ the
instantaneous spatial average of the height, and by $h(x,t)\equiv
H(x,t)-\overline{H(x,t)}$ the relative height, the width of the
interface at time $t$ is given by $W(L,t)\equiv\sqrt{\langle
h^{2}(x,t)\rangle}$ \cite{note-hav}. It is known for a general interface that the width exhibits the Family-Vicsek scaling \cite{Family:1985}, $W(L,t)\sim
L^{\chi}{\cal W}(t/T^\star)$, with the crossover time scale
$T^\star \sim L^z$ defined as the scale over which height fluctuations
spreading laterally correlate the entire interface.
The scaling function ${\cal W}(s)$ behaves as a constant as $s \to \infty$, and
as $s^{\beta}$ as $s\to 0$. Here, $z,\chi,\beta$ are respectively
the dynamic exponent, the roughness exponent, and the growth exponent,
with $z=\chi/\beta$ \cite{Barabasi:1995}. The behavior of ${\cal W}(s)$ encodes the fact that $W(L,t)$ grows with time as $t^\beta$ for $t\ll T^\star$, and 
saturates to an $L$-dependent value $\sim L^\chi$ for $t \gg T^\star$.
For the EW interface in $1d$, one has $z_{{\rm EW}}=2,\chi_{{\rm
EW}}=1/2,\beta_{{\rm EW}}=1/4$.
It is thus evident that for an interface in the thermodynamic limit $L\to\infty$, the width grows
forever with time; indeed, there is no stationary state for the
distribution of fluctuations $h$. For the EW interface in this limit, the $h$-distribution at time $t$, while starting from a flat
interface at $t=0$, is given by a Gaussian with a {\em time-dependent}
variance $W_{{\rm EW}}^{2}(t)\equiv D\sqrt{2/(\pi\nu)}t^{2\beta_{{\rm
EW}}}$ \cite{Barabasi:1995}:
\be
P_{{\rm EW}}(h,t|0,0) =\frac{1}{\sqrt{2\pi W_{{\rm EW}}^{2}(t)}}\exp\left(-\frac{h^{2}}{2W_{{\rm EW}}^{2}(t)}\right).
\label{eq:ew-ht-distr}
\ee
For finite $L$, however, the $h$-distribution at long times $t \gg
T^\star$ is a Gaussian with a {\em time-independent} variance $\sim L^{2\chi}$, corresponding to an
equilibrium stationary state.

Consider the EW interface in the limit $L\to\infty$, and envisage a
dynamical scenario in which the evolution (\ref{eq:ew-ht-distr}) is
repeatedly interrupted with a resetting to the initial flat configuration,
where two successive resets are separated by random time intervals $\tau$ distributed
according to a power-law:
\be
\rho(\tau)=\frac{\alpha}{\tau_0(\tau/\tau_0)^{1+\alpha}};~\tau\in[\tau_0,\infty),~\alpha>0,
\label{eq:ptau}
\ee
where $\tau_0$ is a microscopic cut-off. Note that $\rho(\tau)$
has infinite first and second moments for $\alpha<1$, a finite first
moment for $\alpha>1$, and a finite second moment for $\alpha>2$.

As noted above, in absence of resetting, the
fluctuations grow unbounded in time, and do not have a stationary state. In
this backdrop, we ask: Does introducing resetting lead at long times to
a stationary state with bounded fluctuations? If so, can one
characterize the behavior in the stationary state? How different are these
reset-induced stationary fluctuations from the Gaussian
fluctuations observed in the stationary state for \textit{finite} $L$?
Similar issues were addressed recently in Ref. \cite{Gupta:2014} for fluctuating interfaces being
reset at time intervals distributed as an exponential:
$\rho(\tau)=r\exp(-r\tau)$, in sharp contrast to the power-law that we consider. 
It was shown that a nonzero value of $r$ drives the system to a
nontrivial stationary state that is characterized by non-Gaussian
interface fluctuations, and, in particular, an interface width that is
bounded in time. Changing from an exponential to a power-law is
expected to bring in new effects and surprises, as it happens, e.g.,
even with simple random walks. In the latter case, while a waiting-time
distribution for jumps that is exponential leads to normal diffusion,
changing it to a power-law results in anomalous diffusion with many subtle effects \cite{Klages:2008}.
Thus, we may anticipate new phenomena with a power-law, and indeed do find
them for a general interface including the EW interface.

Our main findings are summarized in Table \ref{table1}. We show that the distribution of interface fluctuations $h$ 
exhibits a rich behavior with multiple crossovers on tuning the exponent $\alpha$ of the power-law distribution (\ref{eq:ptau}). 
For $\alpha>1$, one has at long times a probability distribution for $h$
that no longer spreads in time, but is time independent with power-law
tails; nevertheless, the interface width diverges with time for
$1<\alpha<\alpha^{({\rm w})}$, while a time-independent stationary
behavior emerges only for $\alpha>\alpha^{(\rm w)}$, where $\alpha^{(\rm
w)}\equiv1+2\beta$. By contrast, the dynamics for $0<\alpha<1$ leads at
long times to an $h$-distribution that continually spreads in time, with the interface width growing with time as $t^{\beta}$, similar to the situation in the absence of resetting. Previous
studies for an exponential $\rho(\tau)$ have shown that resetting always leads to a time-independent $h$-distribution with a finite width of the interface \cite{Gupta:2014}, while we here demonstrate that the ensuing scenario is quite different for a power-law $\rho(\tau)$. Besides the two crossovers at $\alpha=1$ and $\alpha=\alpha^{(\rm w)}$, there is another one at $\alpha=\alpha^{(\rm d)}\equiv 1-\beta$, where the time-dependent distribution of fluctuations near the resetting value $h=0$ changes over from a cusp for $0<\alpha<\alpha^{(\rm d)}$ to a divergence for $\alpha^{(\rm d)}<\alpha<1$. This also stands in stark contrast to the case with exponential resetting, where the $h$-distribution at long times always exhibits a cusp singularity \cite{Gupta:2014}. 
\begin{table}[!h]
\centering
\begin{tabular}{|c|c|c|}
\hline 
Inter-reset & & \tabularnewline
time distribution & & \tabularnewline
$\sim\tau^{-(1+\alpha)}$ & $\alpha>1$ & $0<\alpha<1$ \tabularnewline
\hline 
\hline 
Long-time & {\em Stationary}  & {\em Time-dependent} \tabularnewline
distribution & \multicolumn{1}{c|}{$\sim|h|^{-(\alpha+\beta-1)/\beta}$} & $(1/t^{\beta})g_{{\rm r}}(|h|/t^{\beta})$\tabularnewline
of fluctuations            &  ({\em Power-law tails}) & ({\em Scaling form}) \tabularnewline
& & \underline{Around resetting point:} \tabularnewline
& & $\alpha<\alpha^{(\rm d)}$: {\em Cusp} \tabularnewline
& & $\alpha > \alpha^{(\rm d)}$: {\em Divergence} \tabularnewline
& & Cross-over \tabularnewline
& & at $\alpha^{(\rm d)}\equiv1-\beta$\tabularnewline
\hline 
& %
$\alpha<\alpha^{(\rm w)}$: {\em Diverging}& \tabularnewline
Interface width & $\alpha > \alpha^{(\rm w)}$: {\em Stationary} &{\em Diverging} \tabularnewline
& Cross-over at $\alpha^{(\rm w)}\equiv1+2\beta$ &\tabularnewline
\hline 
\end{tabular}
\caption{Summary of long-time behavior of a $1d$ fluctuating interface subject to
stochastic resetting at power-law times. Here, $\beta$ is the growth
exponent characteristic of the universality class for the interface
dynamics in the absence of resetting.}
\label{table1}
\end{table}

The paper is organized as follows. In Section \ref{sec:EW}, we analyze the EW interface for
which remarkably we could derive exact closed-form
expressions for the distribution of fluctuations, and predict thereby a
variety of surprising and subtle effects resulting from resetting. We
corroborate our findings by extensive numerical simulations of a
discrete interface that evolves according to the EW equation.
Our findings for the EW interface are extended to a general interface in
Section \ref{sec:general}, and are checked against numerical simulations
of yet another paradigmatic model of interface evolution, the
Kardar-Parisi-Zhang (KPZ) interface \cite{KPZ:1986}. We draw our conclusions in Section
\ref{sec:conclusions}.
\section{Exact results for the Edwards-Wilkinson interface}
\label{sec:EW}
Here, we compute for the EW interface the quantity $P^{{\rm r}}_{\rm
EW}(h,t|0,0)$, which is the $h$-distribution at
time $t$, while starting from a flat interface at time $t=0$. Let us denote by ${\cal C}\equiv\{h(x,t)\}_{0\leq x \leq L}$ a
configuration of the full interface, with ${\cal C}_0\equiv\{h(x,0)=0\}$ denoting
the initial flat interface. Equation (\ref{eq:eom-ew})
implies Markovian evolution of ${\cal C}$ in the time between successive resets. Then, since each reset defines a renewal of the dynamics, it follows that $P^{{\rm
r}}_{\rm EW}({\cal C},t|{\cal C}_0,0)$, the probability to be in configuration
${\cal C}$ at time $t$ with ${\cal C}={\cal C}_0$ at $t=0$, is given by
the corresponding probability $P_{\rm EW}({\cal C},t|{\cal C}_0,0)$ in the absence of
resetting and the probability $f_{\alpha}(t,t-\tau)$ at time $t$ that the
last reset was at time $t-\tau$ ($\tau\in[0,t]$) by the exact
expression \cite{Nagar:2015}
\be
P^{{\rm r}}_{\rm EW}({\cal C},t|{\cal C}_0,0)=\int_{0}^{t}d\tau~
f_{\alpha}(t,t-\tau)P_{\rm EW}({\cal
C},\tau|{\cal C}_0,0).
\label{eq:pC-time-eqn}
\ee
Integrating over all possible ${\cal C}$'s, noting that $P_{\rm EW}({\cal
C},\tau|{\cal C}_0,0)$ is normalized to unity for every $\tau$, and that
$\int_{0}^{t}d\tau~f_{\alpha}(t,t-\tau)=1$, we check that
$P^{\rm r}_{\rm EW}({\cal C},t|{\cal C}_0,0)$ for every $t$ is also normalized to
unity. The dynamics although Markovian in the full configuration space is not so for the
relative height $h(x,t)$ at a given point $x$ due to the space
derivative of the height field on the right hand side (rhs) of Eq. (\ref{eq:eom-ew}) \cite{Bray:2013}. However, linearity of Eq. (\ref{eq:pC-time-eqn})
allows to get the marginal distribution $P^{{\rm r}}_{\rm EW}(h,t|0,0)$ of the height field $h(x,t)$ by
integrating Eq. (\ref{eq:pC-time-eqn}) over heights $h(y,t)$ at all
positions $y \neq x$ \cite{Gupta:2014}; we get
\be
P^{{\rm r}}_{\rm
EW}(h,t|0,0)=\int_{0}^{t}d\tau~f_{\alpha}(t,t-\tau)P_{\rm EW}(h,\tau|0,0).
\label{eq:pr-time-eqn}
\ee
In terms of the variable $h$, the resetting dynamics we consider corresponds to an
instantaneous jump in its value from $h\ne 0$ to $h=0$, the latter
characterizing the interface at the initial time $t=0$. 

\subsection{Height distribution for $\alpha>1$} For large $t \gg
\tau_0$, it is known that \cite{Godreche:2001,Nagar:2015}
\be
\hspace{-0.5cm}f_{\alpha>1,\tau\ge\tau_{0}}(t,t-\tau)=\frac{1}{\tau_{0}}\Big(\frac{\alpha-1}{\alpha}\Big)\Big(\frac{\tau}{\tau_{0}}\Big)^{-\alpha},\label{eq:fagt1-1}
\ee
and $\int_0^{\tau_{0}}d\tau~f_{\alpha>1,\tau<\tau_{0}}(t,t-\tau)=1-\int_{\tau_{0}}^{t}d\tau~
f_{\alpha>1,\tau\ge\tau_{0}}(t,t-\tau)$. Using Eqs. (\ref{eq:pr-time-eqn}), (\ref{eq:ew-ht-distr}), and the
smallness of $\tau_0$ to write
$\int_{0}^{\tau_0}d\tau~f_{\alpha>1,\tau<\tau_0}(t,t-\tau)P_{\rm EW}(h,\tau|0,0)\approx
P_{\rm EW}(h,\tau_0|0,0)\int_{0}^{\tau_0}d\tau~f_{\alpha>1,\tau<\tau_0}(t,t-\tau)$
give
\bea
&&\hspace{-0.5cm}P_{{\rm EW}}^{{\rm
r},\alpha>1}(h,t|0,0)=\frac{e^{-\frac{z}{\sqrt{\tau_{0}}}}\nu^{1/4}}{(8\pi\tau_0)^{1/4}\sqrt{D}}\left[\frac{\alpha-1}{\alpha}+\frac{1}{\alpha}\Big(\frac{t}{\tau_{0}}\Big)^{1-\alpha}\right]\nonumber \\
 &&\hspace{-0.5cm}+\frac{2\Big(\frac{\alpha-1}{\alpha}\Big)\Big(\frac{z^2}{\tau_{0}}\Big)^{1-\alpha}}{\sqrt{\pi}|h|}\left[\Gamma\left(\beta,\frac{z}{\sqrt{t}}\right)-\Gamma\left(\beta,\frac{z}{\sqrt{\tau_{0}}}\right)\right],\label{eq:ew-agt1-time-depn}
\eea
where $z\equiv h^{2}\sqrt{\pi\nu}/(2^{3/2}D)$, $\beta\equiv 2\alpha-3/2$, while $\Gamma(s,x)$ is the upper incomplete gamma function. 
Expanding the rhs in terms of ordinary Gamma function gives \cite{gamma-note}
\bea
&&\hspace{-0.5cm}P_{{\rm EW}}^{{\rm
r},\alpha>1}(h,t|0,0)=\frac{e^{-\frac{z}{\sqrt{\tau_{0}}}}\nu^{1/4}}{(8\pi\tau_0)^{1/4}\sqrt{D}}\left[\frac{\alpha-1}{\alpha}+\frac{1}{\alpha}\Big(\frac{t}{\tau_{0}}\Big)^{1-\alpha}\right] \nonumber \\
&&\hspace{-0.5cm}+\frac{2\Big(\frac{\alpha-1}{\alpha}\Big)\Big(\frac{z^2}{\tau_{0}}\Big)^{1-\alpha}}{\sqrt{\pi}|h|}\gamma\left(\beta,\frac{z}{\sqrt{\tau_{0}}}\right)\nonumber \\
&&
\hspace{-0.5cm}-\Big(\frac{t}{\tau_{0}}\Big)^{1-\alpha}\Big(\frac{\alpha-1}{\alpha}\Big)\frac{(\frac{2\nu}{t\pi
D^2})^{1/4}}{e^{\frac{z}{\sqrt{t}}}}\sum_{k=0}^{\infty}\frac{\Gamma(\beta)\left(\frac{z}{\sqrt{t}}\right)^{k}}{\Gamma(2\alpha-1/2+k)},
\label{eq:ew-agt1-time-depn-1}
 \eea
where $\gamma(s,x)$ is the lower incomplete gamma function.
On integrating over $h$, and using \cite{Olver:2010} $\int_{0}^{\infty}dz~z^{a-1}\gamma(b,z)=-\Gamma(a+b)/a$
for ${\rm Re}(a)<0$, $\int_{0}^{\infty}dy~y^{2k}\exp(-y^{2})=\Gamma(k+1/2)/2$
for $2k>-1$, and $\sum_{k=0}^{\infty}\Gamma(2\alpha-3/2)\Gamma(k+1/2)/\Gamma(2\alpha-1/2+k)=\sqrt{\pi}/[2(\alpha-1)]$,
we check that $P_{{\rm EW}}^{{\rm
r},\alpha>1}(h,t|0,0)$ is normalized to unity.

While $P_{{\rm EW}}^{{\rm r},\alpha>1}(h,t|0,0)=P_{{\rm EW}}^{{\rm
r},\alpha>1}(-h,t|0,0)$ implies $\langle h \rangle(t)=0~\forall~t$ , the square of the width of the
interface $[W_{{\rm EW}}^{{\rm
r}}(t)]^{2}\equiv\int_{-\infty}^{\infty}dh~h^{2}P_{{\rm EW}}^{{\rm
r}}(h,t|0,0)$ is given by \cite{note-width}
\bea
&&[W_{{\rm EW}}^{{\rm
r},\alpha>1}(t)]^{2}=\left[\frac{\alpha-1}{\alpha}+\frac{1}{\alpha}\Big(\frac{t}{\tau_{0}}\Big)^{1-\alpha}\right]D\sqrt{\frac{2\tau_{0}}{\pi\nu}}\nonumber
\\
&&+\frac{2^{3/2}D\sqrt{\tau_{0}}}{\sqrt{\pi\nu}}\Big(\frac{\alpha-1}{\alpha(2\alpha-3)}\Big)\Big[1-\Big(\frac{t}{\tau_{0}}\Big)^{3/2-\alpha}\Big].
\label{eq:EW-width}
\eea

Equation (\ref{eq:ew-agt1-time-depn-1}) in the limit $t\to\infty$ leads
to a non-trivial stationary state:
\bea
&&\hspace{-0.8cm}P_{{\rm EW,ss}}^{{\rm r},\alpha>1}(h|0)  =\frac{1}{\tau_{0}^{\beta_{{\rm
EW}}}}\mathcal{G}_{{\rm
EW}}\left(\frac{h}{\tau_{0}^{\beta_{{\rm EW}}}}\right); \\
&&\hspace{-0.8cm}\mathcal{G}_{{\rm EW}}(s)
=\left(\frac{\alpha-1}{\alpha}\right)\frac{\nu^{1/4}}{2^{3/4}\pi^{1/4}\sqrt{D}}\exp\left(-\frac{s^{2}\sqrt{\pi\nu}}{2^{3/2}D}\right)\nonumber
\\
&&\hspace{-0.8cm}+\Big(\frac{\alpha-1}{\alpha}\Big)\frac{(\pi\nu/D^2)^{1-\alpha}}{2^{2-3\alpha}\sqrt{\pi}}\frac{1}{|s|^{4\alpha-3}}\gamma\Big(2\alpha-\frac{3}{2},\frac{s^{2}\sqrt{\pi\nu}}{2^{3/2}D}\Big).
\label{eq:ss-EW}
\eea
Moreover, Eq. (\ref{eq:ew-agt1-time-depn-1}) implies a late-time relaxation of the height distribution to
the stationary state as a power-law $\sim1/t^{\alpha-3/4}$. 
Using $\gamma(a,x)/x^{a}\to1/a\,{\rm as}\, x\to0$, and $\gamma(a,x) \to \Gamma(a)$ as $x \to \infty$, we get
\bea
P^{{\rm r},\alpha>1}_{{\rm EW, ss}}(h|0) \sim
\left\{
\begin{array}{ll}
\left(\frac{\alpha-1}{\alpha}\right)\frac{\nu{}^{1/4}}{2^{3/4}(\pi\tau_{0})^{1/4}\sqrt{D}}\Big(\frac{\alpha+1/4}{\alpha-3/4}\Big)
;~h\to 0,\\
\Big(\frac{\pi\nu}{D^{2}\tau_{0}}\Big)^{1-\alpha}\Big(\frac{\alpha-1}{\alpha}\Big)\frac{1}{2^{2-3\alpha}\sqrt{\pi}}\frac{1}{|h|^{4\alpha-3}}\Gamma\Big(2\alpha-\frac{3}{2}\Big);~|h|
\to \infty
\end{array}
        \right. 
\label{eq:a>1limits}
\eea
The stationary state is strongly non-Gaussian with power-law tails $\sim
|h|^{3-4\alpha}$, unlike the Gaussian stationary state for finite $L$.
Also, here one obtains in the distribution a cusp around the resetting
point $h=0$, implying the stationary state to be out of equilibrium
\cite{Evans:2011-1}; this may be contrasted with the equilibrium
stationary state obtained for finite $L$ in the absence of resetting.
Earlier studies on interface resetting at exponential times have shown a
similar phenomenon of a reset-induced non-equilibrium stationary state
\cite{Gupta:2014}. However, in contrast to the power-law case studied
here, the stationary distribution of fluctuations was found to have
stretched exponential tails.

From Eq. (\ref{eq:EW-width}), it follows that for $\alpha>3/2$, the
width at long times relaxes to a stationary value: 
\be
[W_{{\rm EW,ss}}^{{\rm r},\alpha>3/2}]_{{\rm
}}^{2}=\Big(\frac{(\alpha-1)(2\alpha-1)}{\alpha(2\alpha-3)}\Big)D\sqrt{\frac{2\tau_{0}}{\pi\nu}},
\ee
while for $1<\alpha<3/2,$ the width grows indefinitely with time,
behaving at long times as
\be
[W_{{\rm EW}}^{{\rm r},1<\alpha<3/2}(t)]^{2}\approx\frac{2^{3/2}D\sqrt{\tau_{0}}}{\sqrt{\pi\nu}}\Big(\frac{\alpha-1}{\alpha(3-2\alpha)}\Big)\Big(\frac{t}{\tau_{0}}\Big)^{3/2-\alpha}.
\ee
Although the $h$-distribution relaxes to a stationary state for all
$\alpha>1$, it has fat enough tails for $1<\alpha < 3/2$ that the
interface width diverges with time, while a time-independent finite value
results for $\alpha>3/2$. Thus, the interface width exhibits a
crossover at $\alpha^{(\rm w)}=3/2$. This crossover is not observed in
the case of exponential resetting that always yields a finite width of
the interface \cite{Gupta:2014}, and is thus a feature stemming from the
power-law distribution for resetting time intervals.

\subsection{Height distribution for $\alpha<1$}
For large $t \gg \tau_0$, using \cite{Godreche:2001,Nagar:2015} 
\be
f_{\alpha<1}(t,t-\tau)=\frac{\sin(\pi\alpha)}{\pi}\tau^{-\alpha}(t-\tau)^{\alpha-1},
\label{eq:falphalt1}
\ee
and Eq. (\ref{eq:ew-ht-distr}) in Eq. (\ref{eq:pr-time-eqn}) give
\bea
&&P_{{\rm EW}}^{{\rm
r},\alpha<1}(h,t|0,0)=\frac{\nu^{1/4}\Gamma\left(\alpha\right)\sin(\pi
\alpha)}{2^{3/4}\pi^{7/4}\sqrt{D}t^{1/4}} G_{1,3}^{3,0}\left(\begin{array}{c}
\frac{3}{4}\\
\frac{1}{2},0,\frac{3}{4}-\alpha
\end{array}|\frac{z^2}{4t}\right),
\label{eq:ew-a1t1}
\eea
where $G_{p,q}^{\, m,n}\Big(\begin{array}{c}a_{1},\ldots,a_{p}\\
b_{1},\ldots,b_{q}
\end{array}\vert\, z\Big)$ is the Meijer G-function. 
Using \cite{Olver:2010} 
$\int_0^\infty dy~y^{s-1}G_{p,q}^{\, m,n}\Big(\begin{array}{c}a_{1},\ldots,a_{p}\\
b_{1},\ldots,b_{q}
\end{array}\vert\, z\Big)=\prod_{j=1}^m \Gamma(b_j+s)
\prod_{j=1}^n\Gamma(1-a_j-s)/[\prod_{j=m+1}^q
\Gamma(1-b_j-s)\prod_{j=n+1}^p\Gamma(a_j+s)$, we check that $P_{{\rm
EW}}^{{\rm r},\alpha<1}(h,t|0,0)$ is normalized to unity. 

Equation (\ref{eq:ew-a1t1}) suggests the following scaling form of the distribution for
different times:
\be
P_{{\rm EW}}^{{\rm r},\alpha<1}(h,t|0,0)=\frac{1}{t^{\beta_{{\rm
EW}}}}g_{\rm r,{\rm EW}}\Big(\frac{h}{t^{\beta_{{\rm EW}}}}\Big),
\label{eq:ew-alt1-scaling}
\ee
where the scaling function is 
\be
g_{\rm r,{\rm EW}}(s)=\frac{\nu^{1/4}\Gamma\left(\alpha\right)\sin(\pi \alpha)G_{1,3}^{3,0}\left(\begin{array}{c}
\frac{3}{4}\\
\frac{1}{2},0,\frac{3}{4}-\alpha
\end{array}|\frac{s^{4}\nu\pi}{32D^{2}}\right)}{2^{3/4}\pi^{7/4}\sqrt{D}}.
\label{eq:ew-scaling-function-alt1}
\ee
Equation (\ref{eq:ew-alt1-scaling}) implies collapse of the data
for $P^{{\rm r},\alpha<1}_{\rm EW}(h,t|x_0,0)$ at different times on plotting
$t^{\beta_{\rm EW}}P_{{\rm EW}}^{{\rm
r},\alpha<1}(h,t|0,0)$ versus $h/t^{\beta_{\rm EW}}$. 
While the mean $\langle h \rangle$ is zero due to $P_{\rm EW}^{{\rm
r},\alpha<1}(h,t|0,0)$
being even under $h \to -h$, the width grows with time as 
\be
[W_{{\rm EW}}^{{\rm r},\alpha<1}(t)]_{{\rm }}^{2}=Ct^{2\beta_{\rm EW}},
\ee
with $C$ a finite constant:
\be
C \equiv \frac{16 \Gamma(\alpha)\sin(\pi \alpha)D\sqrt{t}}{\pi^{5/2}\sqrt{\nu}}\int_0^\infty dyy^2 G_{1,3}^{3,0}\left(\begin{array}{c}
\frac{3}{4}\\
\frac{1}{2},0,\frac{3}{4}-\alpha
\end{array}|y^{4}\right).
\ee
In fact, all even moments grow with time as $\langle h^{2m}\rangle \sim
t^{2m\beta_{\rm EW}}$, with $m\ge 1$ an integer.

In the limit $t \to \infty$, the rhs of Eq. (\ref{eq:ew-a1t1}) does not
approach a time-independent form. Thus, for $\alpha<1$, the interface
fluctuations do not have a stationary state {\em even in the presence of resetting}. This feature may be contrasted with the
case for $\alpha>1$, where the distribution of fluctuations does relax to a well-defined stationary state
(\ref{eq:ss-EW}) on introducing resetting. Also, while 
exponential resetting of fluctuating interfaces was shown to always lead to a stationary state \cite{Gupta:2014}, our results highlight that such a
scenario does not necessarily hold for resetting at power-law times.

The known small-$x$ and large-$x$ behaviors of $G_{p,q}^{q,0}\left(\begin{array}{c} a_p\\
b_q
\end{array}|x\right)$ yield \cite{asymp-G}
\bea
P^{{\rm r},\alpha<1}_{\rm EW}(h,t|0,0) \sim
\left\{
\begin{array}{ll}
\frac{\nu^{1-\alpha}\Gamma(\alpha-3/4)\Gamma(\alpha-1/4)\sin(\pi
\alpha)}{(D^2t)^{1-\alpha}2^{18/4-5\alpha}\pi^{1+\alpha}|h|^{4\alpha-3}}; h\to 0,~{\rm and}~\frac{3}{4}<\alpha<1,\\
\frac{\nu^{1/4}\Gamma(3/4-\alpha)\Gamma(\alpha)\sin(\pi
\alpha)}{\sqrt{D}t^{1/4}2^{3/4}\Gamma(3/4)\pi^{5/4}}; h\to 0,~{\rm and}~\alpha<\frac{3}{4},\\
\frac{\sqrt{2}\Gamma(\alpha)\sin(\pi\alpha)}{\pi^2}\Big(\frac{\nu \pi}{32D^2t}\Big)^{1/4-\alpha/2}\exp\Big(-\frac{h^2\sqrt{\nu\pi}}{2^{3/2}D^2\sqrt{t}}\Big);|h| \to \infty.
\end{array}
\right. 
\label{eq:a<1limits}
\eea
Thus, on crossing $\alpha=3/4$, the behavior of $P^{{\rm r},\alpha<1}_{\rm
EW}(h,t|0,0) $ as $h\to 0$ crosses over from being with a cusp for
$\alpha<3/4$ to being divergent for $3/4< \alpha<1$. This crossover behavior is explained by
analyzing Eq. (\ref{eq:pr-time-eqn}) in the limit $h \to 0$:
\be
P^{{\rm r},\alpha<1}_{\rm EW}(h \to 0,t|0,0) \sim
\int_{0}^{t}d\tau~\tau^{-\alpha-\beta_{\rm EW}}(t-\tau)^{\alpha-1},
\label{eq:pr-time-eqn-smallh}
\ee
where we have used Eq. (\ref{eq:falphalt1}) and the fact that $P_{\rm
EW}(h\to 0,\tau>0|0,0)$ = a finite constant \cite{note-Phto0}. The
integral on the rhs is finite for $\alpha+\beta_{\rm EW}<1$, whereby it contributes a cusp, and is divergent for $\alpha+\beta_{\rm
EW} \ge 1$. A crossover in behavior is then expected at
$\alpha^{(\rm d)}=1-\beta_{\rm EW}=3/4$. For a general interface with growth
exponent $\beta$, we predict a similar crossover from cusp to
divergence in the $h$-distribution close to the
resetting location at $\alpha^{(\rm d)} \equiv 1-\beta$. 

Similar to the $h$-distribution
(\ref{eq:ew-ht-distr}) for the EW interface in $1d$,  a single diffusing
particle has a spatial distribution that is Gaussian, the difference
being that the variance grows with time as $t^{2\beta_{\rm
EW}}$ in the former and as $t^{2\beta_{\rm Diff}}$ in the latter, with $\beta_{\rm Diff}=1/2$. When subject to resetting at power-law
times, we may on the basis of the above discussion predict the
spatial distribution of the diffusing particle
close to the resetting location to exhibit a crossover from cusp to
divergence at $\alpha^{(\rm d)}=1-\beta_{\rm Diff}=1/2$; this is
indeed borne out by our exact results in Ref. \cite{Nagar:2015}.

\subsection{Numerical simulations}
\begin{figure}[!h]
\centering
\includegraphics[width=8cm]{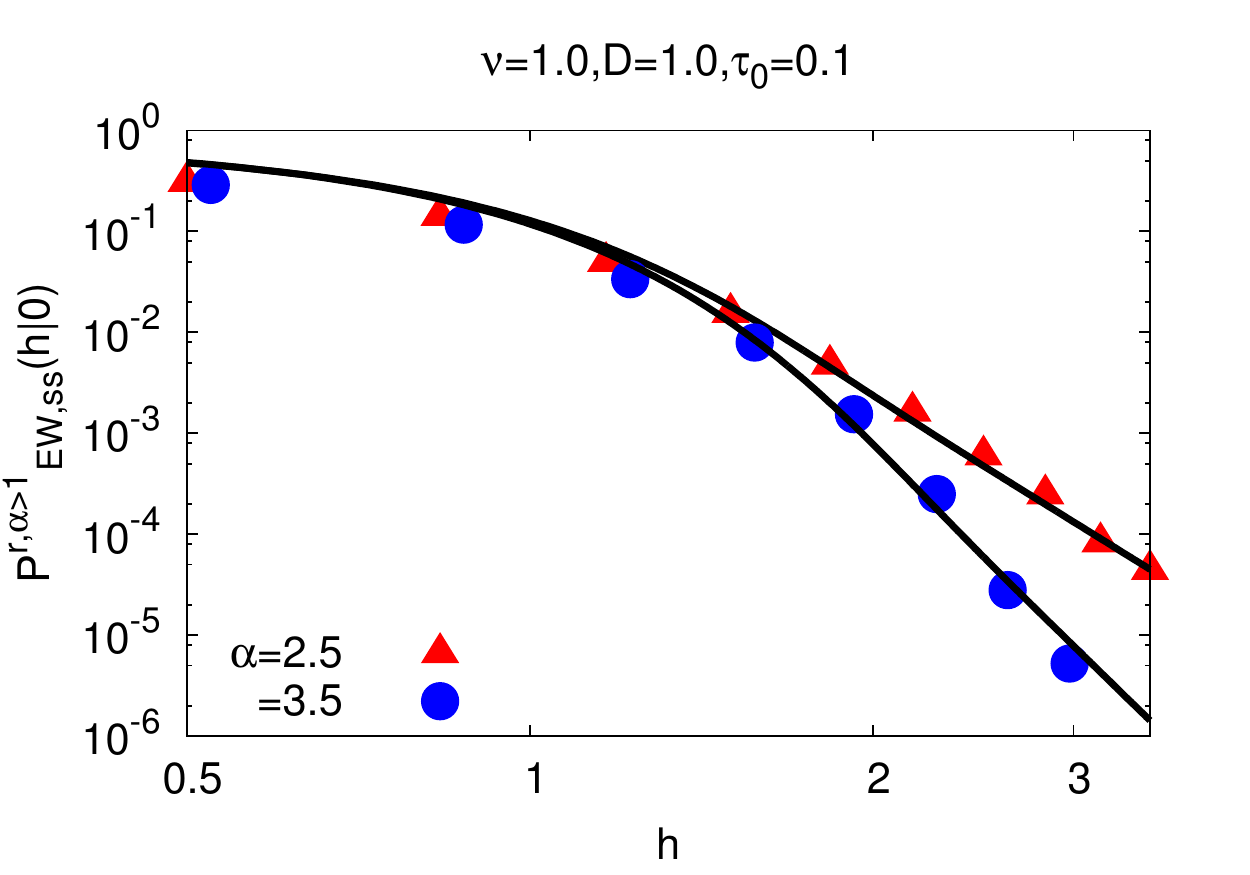}
\caption{Steady-state height distribution for the $1d$ EW interface subject
to resetting at power-law times, with $\alpha>1.$ The points refer
to numerical simulations of a discrete interface of size $L=2^{14}$, while lines refer to the exact result (\ref{eq:ss-EW}). The various parameter
values are indicated in the figure.}
\label{fig:ew-ss-agt1}
\end{figure}

\begin{figure}[!h]
\centering
\includegraphics[width=15cm]{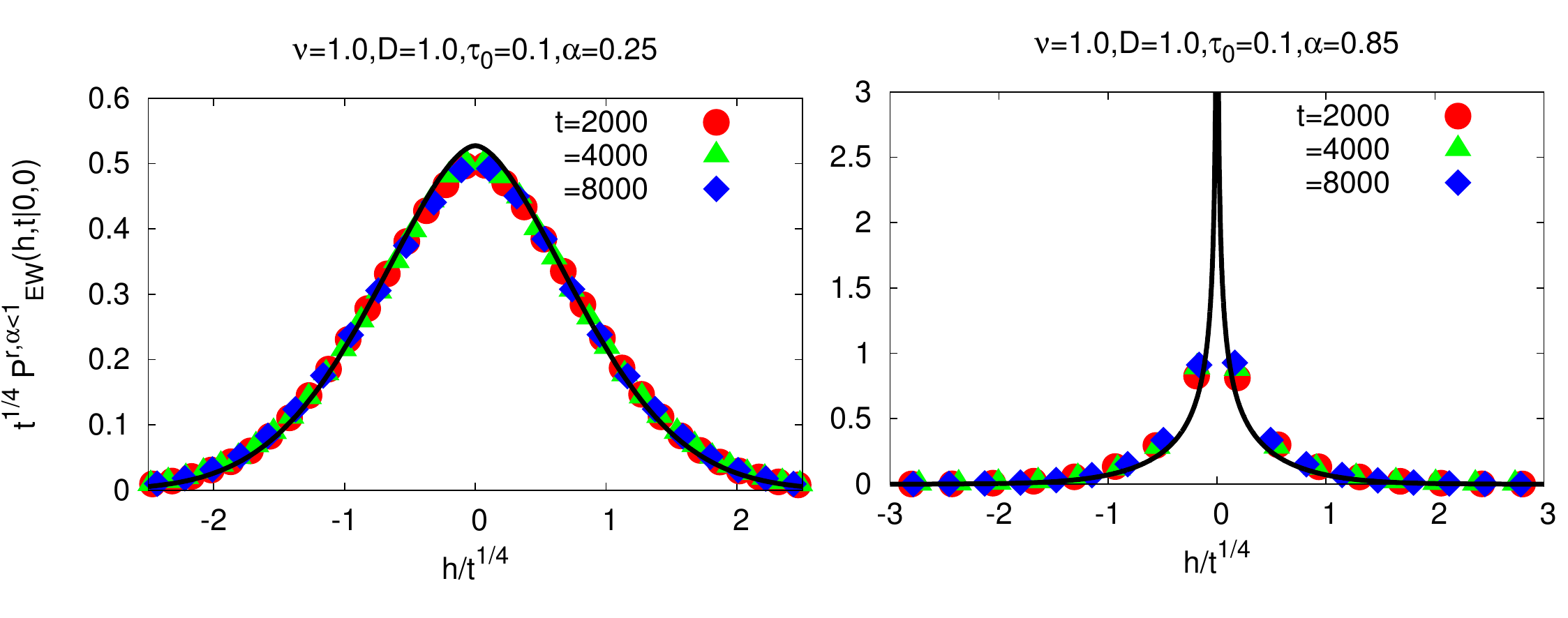}
\caption{Time-dependent height distribution for the $1d$ EW interface subject
to resetting at power-law times, with $\alpha<1.$ The data points are obtained
from numerical simulations of a discrete interface of size $L=2^{14}$. Collapse of the data
for different times follows the scaling form (\ref{eq:ew-alt1-scaling}),
with $\beta_{\rm EW}=1/4$. Here, the lines denote the exact scaling function (\ref{eq:ew-scaling-function-alt1}).}
\label{fig:ew-time-alt1}
\end{figure}

To confirm our results for the EW interface, we now
report on numerical simulations performed on a discrete
$1d$ periodic interface $\{H_{i}(t)\}_{i=1,2,\ldots,L}$ that evolves
at times $t_{n}=n\Delta t,$ with $n$ an integer and time step $\Delta t\ll1$.
We start with a flat interface, $H_{i}(0)=0~\forall~i$, and its evolution
according to the EW dynamics is interrupted by a reset to the initial
configuration, with two successive resets separated by time intervals $\tau$ sampled from the distribution
(\ref{eq:ptau}). Figure \ref{fig:ew-ss-agt1} shows for two
representative values of $\alpha>1$ a comparison of the simulation
results with the stationary-state distribution (\ref{eq:ss-EW}). One may observe a very good
agreement between theory (lines) and simulation (points). Figure
\ref{fig:ew-time-alt1} shows for two
representative values of $\alpha<1$ a collapse of the simulation data
for different times in accordance with the scaling form (\ref{eq:ew-alt1-scaling}), with the lines
showing the exact scaling function (\ref{eq:ew-scaling-function-alt1}).
One may note on crossing $\alpha=3/4$ the change in the behavior of
$P^{{\rm r},\alpha<1}_{\rm
EW}(h,t|0,0) $ as $h\to 0$, from being with a cusp to being divergent,
as predicted by our exact results. 

\section{Predictions for a general interface}
\label{sec:general}
Consider now a general interface characterized by scaling exponents $\chi,z$, and
$\beta=\chi/z$, for which the distribution $P(h,\tau|0,0)$
in the limit $\tau\to\infty,|h|\to\infty$, keeping $|h|/\tau^{\beta}$
fixed and finite, has the scaling form $P(h,\tau|0,0)\sim(1/\tau^{\beta})g\left(h/\tau^{\beta}\right)$, where
$g(s)=g(-s)$ is the scaling function. Normalization requires $P(|h| \to
\infty,\tau|0,0)\to 0 ~\forall~\tau$. Provided the interface dynamics is
Markovian in the full configuration space, Eq. (\ref{eq:pr-time-eqn})
will still hold for such an interface. 

\subsection{Height distribution for $\alpha>1$}
Using Eq. (\ref{eq:pr-time-eqn}) and the expression for $f_{\alpha>1}(t,t-\tau)$, we get 
\bea
&& P^{{\rm
r},\alpha>1}(h,t|0,0)=P(h,\tau_{0}|0,0)\left[\frac{\alpha-1}{\alpha}+\frac{1}{\alpha}\Big(\frac{t}{\tau_{0}}\Big)^{1-\alpha}\right]\nonumber
\\
&&+\int_{\tau_{0}}^td\tau~\frac{1}{\,\tau_{0}}\Big(\frac{\alpha-1}{\alpha}\Big)\Big(\frac{\tau}{\tau_{0}}\Big)^{-\alpha}P(h,\tau|0,0).
\eea
The stationary state, obtained in the limit $t\to \infty$, has the large-$h$
behavior given by $P_{{\rm ss}}^{{\rm r},\alpha>1}(|h|\to
\infty|0)\sim\int_{\tau^{\star}}^{\infty}d\tau~(1/\tau^{\alpha+\beta})g(|h|/\tau^{\beta})$, where $\tau^{\star}$ is such that the scaling form
for $P(h,\tau|0,0)$ holds for $\tau>\tau^{\star}$, and we have used the
smallness of $\tau_0$ to neglect the contribution of $P(h,\tau_0|0,0)$
as $|h| \to \infty$. The above equation
implies a power-law decay of the height distribution at
the tails, 
\be
P_{{\rm ss}}^{{\rm r},\alpha>1}(|h|\to
\infty|0)\sim\frac{1}{|h|^{(\alpha+\beta-1)/\beta}},\label{eq:ph-generic-scaling-alt1}
\ee
which therefore predicts a crossover in the interface width, from finite
to infinite, at $\alpha^{(\rm w)}=1+2\beta$. These predictions match
with our exact results for the EW interface.

\subsection{Height distribution for $\alpha<1$}
Using Eqs. (\ref{eq:pr-time-eqn})
and (\ref{eq:falphalt1}), we get 
\be
P^{{\rm
r},\alpha<1}(h,t|0,0)=\int_{0}^td\tau~\tau^{-\alpha}(t-\tau)^{\alpha-1}P(h,\tau|0,0),
\ee
so that for large $t$, one has the large-$h$ behavior $P^{{\rm
r},\alpha<1}(|h|\to
\infty,t\to
\infty|0,0)\sim\int_{\tau^{\star}}^td\tau~\tau^{-(\alpha+\beta)}(t-\tau)^{\alpha-1}g(|h|/\tau^{\beta})$, where $\tau^{\star}$ is such that the scaling form
for $P(h,\tau|0,0)$ holds for $\tau>\tau^{\star}$. One then arrives at the scaling form 
\be
P^{{\rm r},\alpha<1}(h,t|0,0)\sim\frac{1}{t^\beta}g_{\rm r}\Big(\frac{h}{t^\beta}\Big),
\label{eq:general-alt1-scaling}
\ee
consistent with the result (\ref{eq:ew-alt1-scaling}) obtained for the
EW interface. In the paragraph following Eq. (\ref{eq:pr-time-eqn-smallh}), we have already discussed that close to
the resetting location, $P^{{\rm r},\alpha<1}(h,t|0,0)$ is expected to exhibit a
crossover in behavior, from cusp to divergence, across $\alpha^{(\rm d)}=1-\beta$.

\subsection{Numerical simulations for a KPZ interface}
\begin{figure}[!h]
\centering
\includegraphics[width=8cm]{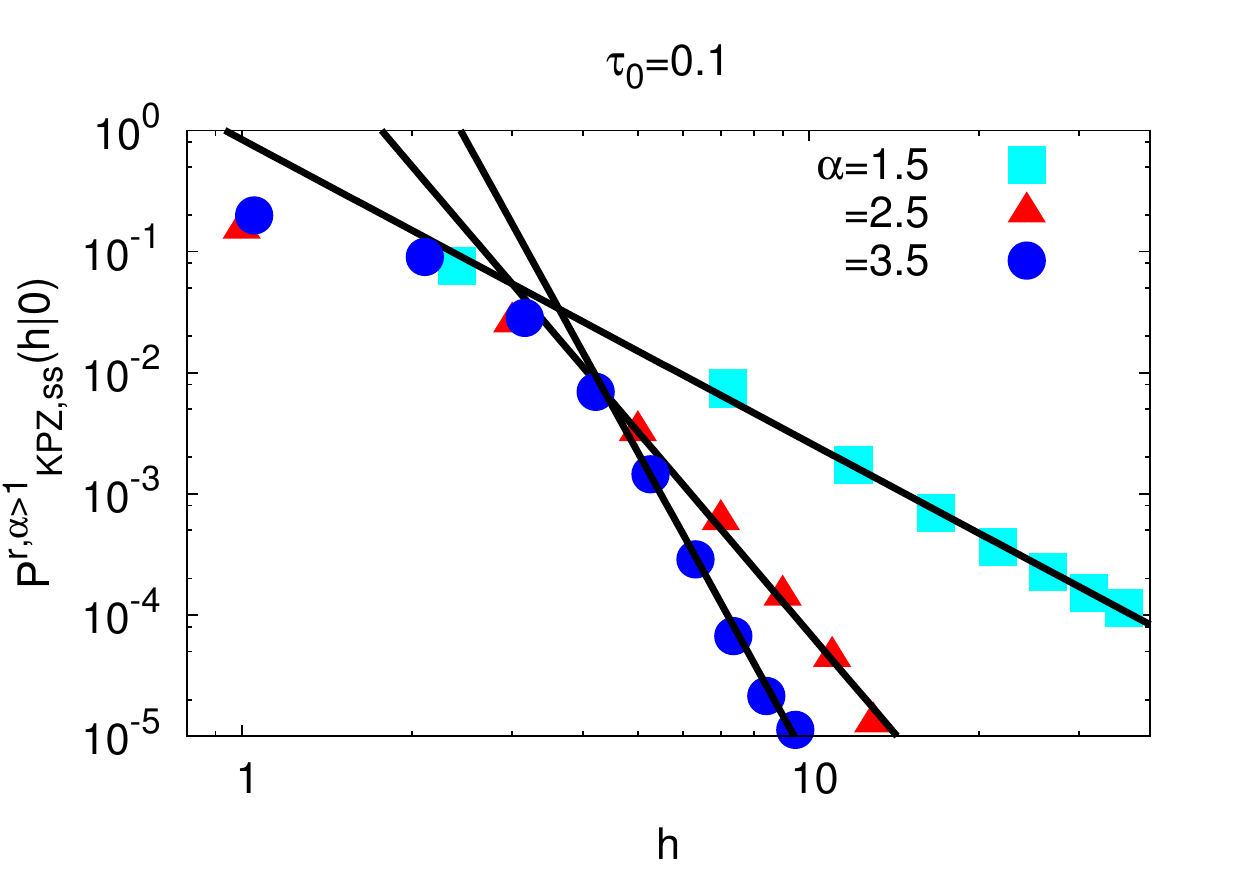}
\caption{Steady-state height distribution for the $1d$ KPZ interface subject
to resetting at power-law times, with $\alpha>1.$ The points refer
to numerical simulations of a discrete interface of size $L=2^{14}$, while lines refer to the predicted power-law tails, Eq. (\ref{eq:ph-generic-scaling-alt1}).}
\label{fig:kpz-ss-agt1}
\end{figure}

\begin{figure}
\centering
\includegraphics[width=15cm]{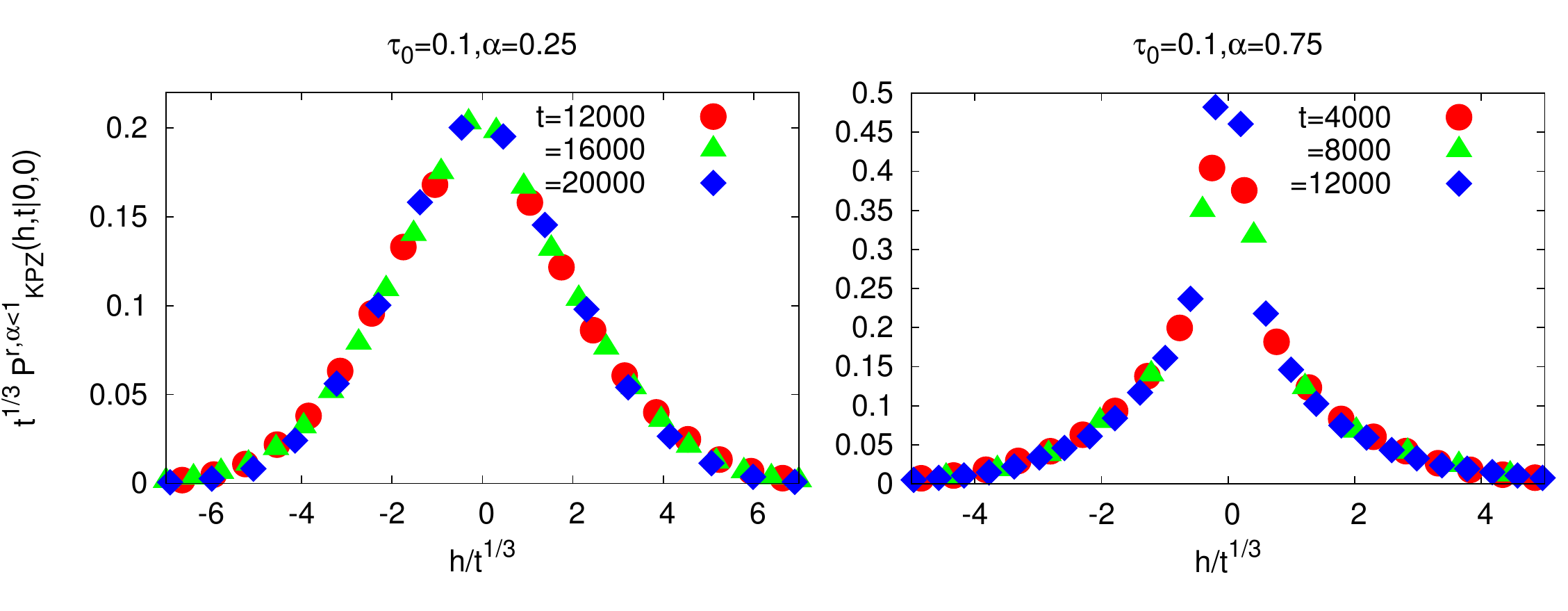}
\caption{Time-dependent height distribution for the $1d$ KPZ interface subject
to resetting at power-law times, with $\alpha<1.$ The data points are obtained
from numerical simulations of a discrete interface of size $L=2^{14}$. Collapse of the data
for different times follows the scaling (\ref{eq:general-alt1-scaling}) with $\beta=\beta_{\rm KPZ}=1/3$.}
\label{fig:kpz-time-alt1}
\end{figure}

Our derived predictions for the different behaviors of fluctuations are
summarized in Table \ref{table1}; to confirm
their validity beyond the EW interface, we now consider a $1d$ periodic interface in the KPZ universality class. In this case, the evolution
equation (\ref{eq:eom-ew}) is augmented by a non-linear term:
\be
\frac{\partial H}{\partial t}=\nu\frac{\partial^{2}H}{\partial
x^{2}}+\frac{\lambda}{2}\Big(\frac{\partial H}{\partial
x}\Big)^{2}+\eta(x,t),
\label{eq:eom-kpz}
\ee
and the exponents $z,\chi,\beta$ have the values $z_{{\rm
KPZ}}=3/2,\chi_{{\rm KPZ}}=1/2,\beta_{{\rm KPZ}}=1/3$.
To check our predictions, we performed numerical simulations of a
discrete $1d$ periodic interface $\{H_i(t)\}_{1,2,\ldots,L}$, which evolves in discrete times $t$ according to the following dynamics of the ballistic deposition
model in the KPZ universality class \cite{Barabasi:1995,Halpin-Healy:1995,Krug:1997},
\be
H_i(t+1)={\rm max}[H_{i-1}(t),H_i(t)+1,H_{i+1}(t)],
\label{eq:kpz-discrete-evolution}
\ee
and is additionally reset to the initial flat configuration
$H_i(0)=0~\forall~i$. As in all
our discussions in this paper, we take two successive resets to be
separated by a random interval $\tau$ sampled from
the power-law distribution (\ref{eq:ptau}). The results of numerical
simulations shown in Figs. \ref{fig:kpz-ss-agt1} and \ref{fig:kpz-time-alt1} are fully consistent with the predictions in Table
\ref{table1}.

\section{Conclusions}
\label{sec:conclusions}
In this work, we studied the problem of a fluctuating interface in one
dimension, whose dynamics is interrupted with resetting
to a given initial configuration, namely, a flat configuration, after random time intervals $\tau$
distributed as a power-law $\sim \tau^{-(1+\alpha)};~\alpha > 0$. With
respect to an earlier study that considered resetting at exponential
times \cite{Gupta:2014}, here we demonstrated by exact analytical results and numerical
simulations that the dynamics exhibits new and interesting reset-induced effects, including many crossover
phenomena. We found that the distribution of interface fluctuations $h$
is time dependent for $0 < \alpha<1$, and, moreover, the distribution around the resetting value $h=0$ shows a crossover from a cusp for $\alpha< \alpha^{({\rm d})}$ to a divergence for $\alpha^{({\rm d})}<\alpha<1$, where $\alpha^{({\rm d})}\equiv1-\beta$. The interface width grows with time as $t^{\beta}$, which is also the situation in the absence of resetting. For $\alpha>1$, by contrast, the distribution of fluctuations is time independent, yet the interface width diverges with time for $1 < \alpha < \alpha^{({\rm w})}$, but is time independent with a finite value for $\alpha > \alpha^{({\rm w})}$, with $\alpha^{({\rm w})}\equiv1+2\beta$.
These features may be contrasted with resetting at exponentially-distributed times that always leads to a time-independent
state at long times with a finite value of the interface width \cite{Gupta:2014}.

The qualitative behaviors and the cross-overs observed here for the
many-body interacting system of an interface were previously reported by
us to be present also for a single particle undergoing stochastic
resetting at power-law times \cite{Gupta:2014}, and their origin may be
traced to the form of the probability $f_\alpha(t,t-\tau)$,
Eqs. (\ref{eq:fagt1-1}) and (\ref{eq:falphalt1}), a vital
ingredient in determining the behavior of fluctuations for either
system, and, thence, essentially to the properties of the resetting time distribution $\rho(\tau)$, Eq. (\ref{eq:ptau}). Indeed, for $0 < \alpha < 1$, the presence of fluctuations that are unbounded in time (so that there is no stationary state even at long times) is due to the fact that for this range of $\alpha$, the average gap $\langle \tau \rangle$ between successive resets is infinite, so that there are only a small number of reset events in a given time, and in between the fluctuations may grow unbounded in time. The situation is very different for $\alpha>1$; in this case, a finite $\langle \tau \rangle$ implies frequent resets in a given time, so that the fluctuations cannot grow unbounded in time, and hence, the dynamics exhibits a long-time stationary state. This physical picture was proposed and validated by us in the context of a single diffusing particle in Ref. \cite{Nagar:2015}, and the present work demonstrates its generality beyond a single particle to a many-interacting particle system. Besides the example of a fluctuating interface, it would be
interesting to study if and how resetting leads to new behaviors in
other many-particle interacting systems, such as models of
interacting particles diffusing on a lattice.



\begin{thebibliography}{99}
\bibitem{note-hav}For a translationally invariant interface, one has $\langle h \rangle=0$.
\bibitem{asymp-G}In the limit $|x|\to\infty$, it is known that \cite{Luke:1969} $G_{p,q}^{q,0}(\begin{array}{c}
a_p\\
b_q
\end{array}|x)
 \approx (2\pi)^{(\sigma-1)/2}/\sqrt{\sigma}\exp(-\sigma
 x^{1/\sigma})x^\theta$, with $|{\rm arg}~x|\le
 (\sigma+\epsilon)\pi-\delta$, $\delta>0$, and $\sigma\equiv q-p$, where
 $\epsilon=1/2$ if $\sigma=1$ and $\epsilon=1$ if $\sigma>1$, while
 $\sigma\theta=(1/2)(1-\sigma)+\sum_{h=1}^q b_h-\sum_{h=1}^p a_h$. On the other hand, as $x\to 0$, we have
 \cite{mathematica} $
x^{-a} \Big(\Big(\frac{\sqrt{\pi } \Gamma(\frac{3}{4}-a)}{\Gamma \left(\frac{3}{4}\right)}-\frac{2 \left(\sqrt{\pi } \Gamma
   \left(\frac{1}{4}-a\right)\right) \sqrt{x}}{\Gamma \left(\frac{1}{4}\right)}+\frac{2 \sqrt{\pi } \Gamma \left(\frac{3}{4}-a\right) x}{(4 a+1) \Gamma
   \left(\frac{3}{4}\right)}+O\left(x^{3/2}\right)\Big) x^a+\left(\frac{\Gamma \left(a-\frac{3}{4}\right) \Gamma \left(a-\frac{1}{4}\right)
   x^{3/4}}{\Gamma (a)}+O\left(x^{5/4}\right)\right)\Big)$.
\bibitem{note-Phto0}For a general interface also, we have $P(h\to
0,\tau>0|0,0)$ = a finite constant. Indeed, even on starting with a flat
interface, so that $P(h,0|0,0)=\delta(h)$, the dynamical evolution will
roughen the interface in a finite time, so that $P(h,\tau>0|0,0)$ will
have a finite width, and in particular, will result in $P(h\to
0,\tau>0|0,0)$ = a finite constant.
\end{thebibliography}
\end{document}